\documentclass[aps,groupedaddress,fleqn,nofootinbib,book,longbibliography]{revtex4}
\usepackage{graphicx}
\usepackage{xcolor}
\usepackage{amsmath,amssymb}
\usepackage{float}
\usepackage{physics}
\usepackage{amsfonts}
\usepackage{verbatim}
\usepackage{flushend}
\usepackage{balance}
\usepackage{endnotes}
\usepackage{footnote}
\usepackage{adjustbox}
\usepackage{gensymb}
\usepackage{float}
\usepackage{epigraph}
\usepackage{xcolor}
\usepackage[utf8]{inputenc}
\usepackage{setspace}
\usepackage[utf8]{inputenc}
\usepackage[T1]{fontenc}
\usepackage{comment}
\usepackage{rotating}
\usepackage{floatpag}
\rotfloatpagestyle{empty}
\usepackage{makeidx}
\usepackage{caption}
\usepackage{arydshln}
\usepackage{booktabs}
\makeindex

\newcommand{\beq}{\begin{eqnarray}}
\newcommand{\eeq}{\end{eqnarray}}
\usepackage{amsmath}

\begin{document}

\title{\LARGE{Theory of liquids}\\
\large{From Excitations to Thermodynamics\footnote[1]{Cambridge University Press, 2023}}
}
\author{Kostya Trachenko}
\address{School of Physical and Chemical Sciences, Queen Mary University of London, Mile End Road, London, E1 4NS, UK}

\begin{abstract}
Of the three basic states of matter, liquid is perhaps the most complex. While its flow properties are described by fluid mechanics, its thermodynamic properties are
often neglected, and for many years it was widely believed that a general theory of liquid thermodynamics was unattainable. In recent decades that view has been
challenged, as new advances have finally enabled us to understand and describe the thermodynamic properties of liquids. This book explains the recent developments
in theory, experiment and modelling that have enabled us to understand the behaviour of excitations in liquids and the impact of this behaviour on heat capacity and
other basic properties. Presented in plain language with a focus on real liquids and their experimental properties, this book is a useful reference text for researchers
and graduate students in condensed matter physics and chemistry, as well as for advanced courses covering the theory of liquids.
\end{abstract}

\maketitle
\newpage

{\scalebox{0.25}{\includegraphics{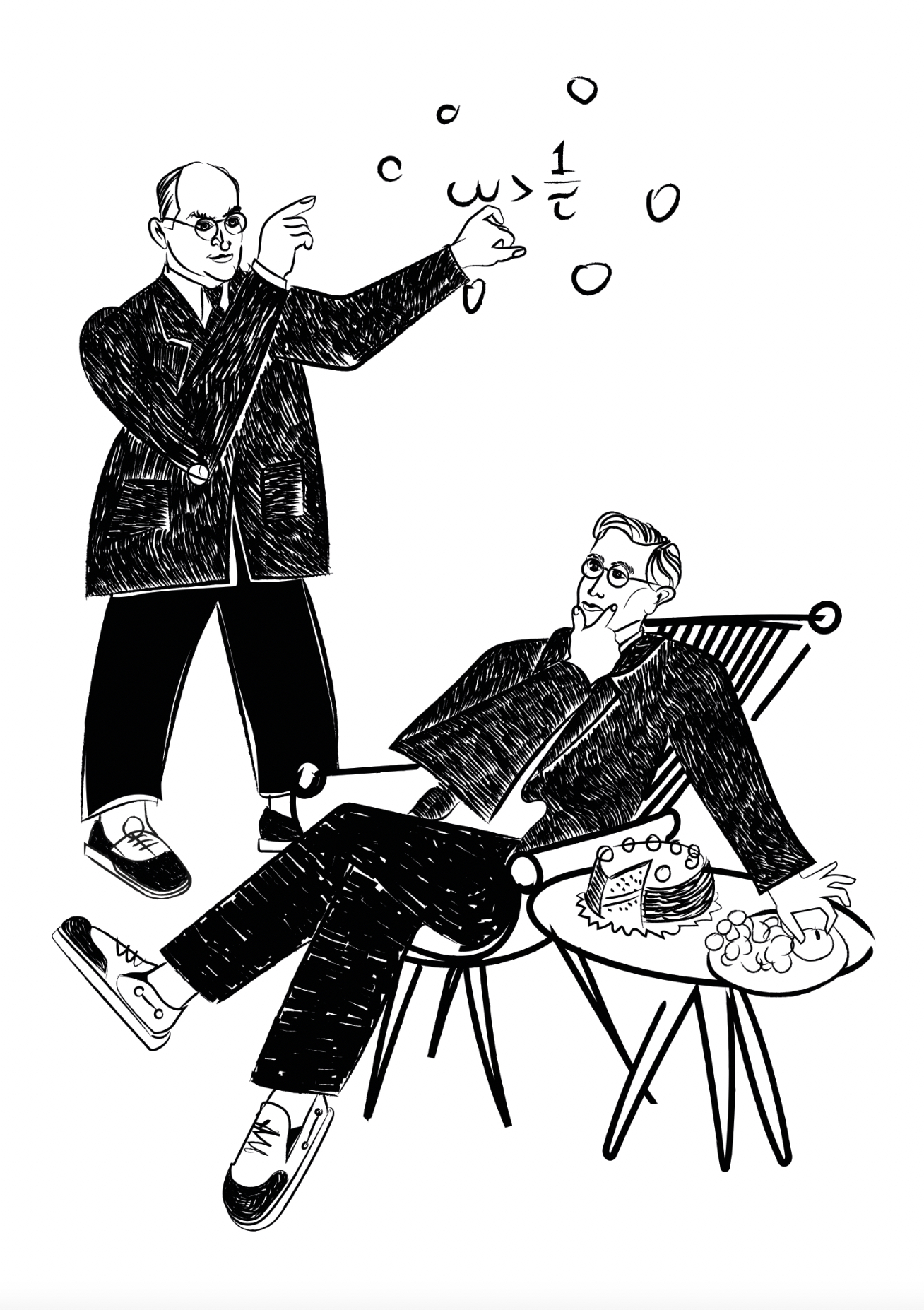}}}\\

Frenkel discusses his theory of liquids with Mott. Based on Mott's recollection in Section 8b.\\
Illustrated by Ivan Shkoropad.

\newpage

\begin{center}
{\bf Contents}\\
\end{center}

{\bf 1. Preface}\\

{\bf 2. Introduction}\\

{\bf 3. Experimental heat capacity}\\

{\bf 4. Excitations in solids and gases}\\
\indent A. Harmonic solids and quadratic forms\\
\indent B. Continuum approximation: elasticity and hydrodynamics\\
\indent C. Excitations in quantum liquids\\
\indent D. Intermediate summary\\

{\bf 5. Energy and heat capacity of solids}\\
\indent A. Harmonic solids\\
\indent B. Weak anharmonicity\\

{\bf 6. First-principles description of liquids: exponential complexity}\\

{\bf 7. Liquid energy and heat capacity from interactions and correlation functions}\\
\indent A. The method\\
\indent B. The problem\\
\indent C. A better alternative?\\
\indent D. Structure and thermodynamics\\

{\bf 8. Frenkel theory and Frenkel book}\\
\indent A. Liquid relaxation time\\
\indent B. Collective excitations in the Frenkel theory\\
\indent C. Continuity of liquid and solid states and Frenkel-Landau argument: first ignition of the glass transition debate\\

{\bf 9. Collective excitations in liquids post-Frenkel}\\
\indent A. The telegraph equation and gapped excitations in liquids\\
\indent B. Another solution of the telegraph equation\\
\indent C. The first glimpse into liquid thermodynamics and collective modes\\
\indent D. Symmetry of liquids: viscoelasticity and elastoviscosity\\
\indent E. Collective modes in generalised hydrodynamics\\
\indent F. Experimental evidence of collective excitations in liquids\\
\indent G. Decay of excitations in liquids and solids\\

{\bf 10. Molecular dynamics simulations}\\

{\bf 11. Liquid energy and heat capacity in the liquid theory based on collective excitations}\\
\indent A. Liquid energy and heat capacity\\
\indent B. Comparison to the integral over interaction and correlation functions\\
\indent C. Anharmonic effects\\
\indent D. Quantum effects\\
\indent E. Verification of the liquid theory based on collective excitations\\
\indent F. Molecular liquids and quantum excitations\\

{\bf 12. Quantum liquids: excitations and thermodynamic properties}\\
\indent A. Liquid 4He\\
\indent B. Liquid 3He\\
\indent C. Quantum indistinguishability\\

{\bf 13. Sui Generis}\\

{\bf 14. Connection between phonons and liquid thermodynamics: a historical survey}\\
\indent A. Early period\\
\indent B. Late period\\

{\bf 15. The supercritical state}\\
\indent A. What happens at high temperature?\\
\indent B. The supercritical state and its transition lines \\
\indent C. Applications of supercritical fluids \\
\indent D. C transition \\
\indent E. Excitations and thermodynamic properties \\
\indent F. Liquid elasticity length and particle mean free path \\
\indent G. Heat capacity of matter: from solids to liquids to gases \\

{\bf 16. Minimal quantum viscosity from fundamental physical constants}\\
\indent A. Viscosity minima and dynamical crossover\\
\indent B. Minimal quantum viscosity and fundamental constants \\
\indent C. Elementary viscosity, diffusion constant and uncertainty principle \\
\indent D. Viscosity minima in quantum liquids \\
\indent E. Lower bound of thermal diffusivity \\
\indent F. Practical implications \\
\indent G. The Purcell question: why do all viscosities stop at the same place? \\
\indent H. Fundamental constants, quantumness and life \\
\indent I. Another layer to the anthropic principle\\

{\bf 17. Viscous liquids }\\
\indent A. Collective excitations, energy and heat capacity\\
\indent B. Entropy \\
\indent C. Heat capacity at the liquid-glass transition \\
\indent D. Glass transition revisited \\

{\bf 18. A short note on theory}\\

{\bf 19. Excitations and thermodynamics in other disordered media: spin waves and spin glass systems}\\

{\bf 20. Evolution of excitations in strongly-coupled systems}\\

{\bf References}\\

{\bf Index }\\

\newpage

\newpage

\section{Preface}

In 1946, Born and Green wrote \cite{borngreen}:

\begin{quotation}
``It has been said that there exists no general theory of liquids.''
\end{quotation}

About 60 years later, Granato observed \cite{granato} that nothing is said about liquid specific heat \index{specific heat} in standard introductory textbooks, and little or nothing in advanced texts, adding that there is little general awareness of what the basic experimental facts to be explained are. This was reflected in his teaching: Granato recalls how he enjoyed teaching theories of specific heat of gases and solids but lived in constant fear that a student might ask ``how about liquids''? That question was never asked in lectures over many years. Granato attributes this to an issue with our teaching practice where unsolved problems in physics are not sufficiently covered.

Problems involved in liquid theory are fundamental. These problems are well illustrated by two related assertions. These assertions are important to quote in full. The first quote is from the Statistical Physics textbook by Landau and Lifshitz (LL) \index{Landau and Lifshitz} \cite{landaustat}. LL discuss general thermodynamic properties of liquids and explain why they {\it can not} be calculated, contrary to solids and gases. Section 66 opens with a forceful statement of the problem:

\begin{quotation}
``Unlike gases and solids, liquids do not allow a calculation in a general form of the thermodynamic quantities or even of their dependence on temperature. The reason lies in the existence of a strong interaction \index{strong interactions} between the molecules of the liquid while at the same time we do not have the smallness of the vibrations which makes the thermal motion in solids especially simple. The strength of the interaction between molecules makes it necessary to know the precise law of interaction in order to calculate the thermodynamic quantities, and this law is different for different liquids.''
\end{quotation}

LL emphasize this point later in the opening of Section 74 discussing the van der Waals equation.

The second quote is from Pitaevskii \cite{Pitaevskii} \index{Pitaevskii}:

\begin{quotation}
``Differently from a gas, the interaction terms in the equation of state for liquids are not small. Therefore liquid properties depend strongly on the specific character of interaction between molecules. Liquid theory generally lacks a small parameter \index{small parameter} which could be used to simplify a theory. It is impossible to obtain any analytical formulas for thermodynamic properties of liquids.''
\end{quotation}

The last sentence in the second quote is consistent with the view held by Landau. According to Peierls \cite{peierls-frenkel}, Landau has always maintained that a theory of liquids is impossible.

The above problems seemingly seal the fate of liquid theory and preclude its development at the level nowhere near close to well-advanced theories of the two other states of matter, solids and gases, which we have been enjoying for a long time.

Appreciating the problems in the quotes above, I wondered whether these problems are as arresting as they sound. Are liquids really so complex that we can not hope for a theory explaining their most basic properties and can not teach liquids to our students? Could it be that we became too attached to ideas used in theories of solids and gases and did not manage to move on with liquids using a different approach? Would this approach require a modification of solid state and gas theories or do we need to come up with something entirely different in order to understand liquids?

It is important to ask at this point why the problems outlined by Landau, Lifshitz and Pitaevskii do not emerge in the solid state theory where interactions are also strong and system-specific as in liquids? An important premise of statistical physics is that an interacting system is governed by its excitations \cite{landaustat,landaustat1}. In solids, these are collective excitations, phonons. Understanding phonons and relating them to solid properties is at the heart of the remarkable success of the solid state theory developed over 100 years ago. Why can't we use the same approach to liquids based on phonons? One quick answer could be that liquids flow and hence do not have fixed reference points which we can use to perform the harmonic expansion and derive these phonons. This is not quite the case: liquid particles oscillate around quasi-equilibrium points at time shorter than liquid relaxation time before jumping to new positions. Discussing the theory of this motion and its implications for phonons in liquids and liquid thermodynamic properties constitutes a large part of this book.

The second important observation is related to the predominant approach adopted in liquid theories in the last century. The first attempt at liquid theory using phonons was undertaken by Sommerfeld and Brillouin over 100 years ago. This was around the same time when the papers by Einstein and Debye were published and laid the foundations of the modern solid state theory based on phonons. Developing this line of inquiry in liquids had largely stopped soon after, and theories of liquids and solids diverged at the point of a fundamental approach. Whereas the solid state theory continued to be developed on the basis of phonons, theories of liquids started to use the approach based on interatomic interactions and correlation functions. As discussed in this book in detail, this approach faced several inherent limitations and fundamental problems. This was largely because the problems stated by Landau, Lifshitz and Pitaevskii are inevitable in this approach. From the perspective of this book, the important problem is that the approach based on interatomic interactions and correlation functions has not resulted in a physical understanding of the most basic experimental properties of real liquids such as specific heat.

A set of new results have emerged in the last few decades related to collective excitations in liquids. It has taken a combination of experiment, theory and modelling to understand phonons in liquids well enough to connect them to liquid thermodynamic properties. Recall that this connection between phonons and thermodynamic properties was the basis of the Einstein and Debye approach to solids which laid the foundations of the modern solid state theory. The upshot is that the liquid theory can still be constructed on the basis of phonons, but the key point is that, differently from solids, the phase space available to phonons in liquids is not fixed but is {\it variable} instead. In particular, this phase space reduces with temperature. This reduction quantitatively explains the experimental liquid data and in particular the reduction of liquid specific heat from the solidlike value of $c_v=3k_{\rm B}$ to the ideal gas value $c_v=\frac{3}{2}k_{\rm B}$ with temperature. It has taken another decade to obtain an independent verification of this theory.

The small parameter in the liquid theory is therefore the same as in the solid state theory: small phonon displacements. However, in important difference to solids, this small parameter operates in a variable phase space. This addresses the problems stated by Landau, Lifshitz and Pitaevskii above.

We can see why earlier liquid theories diverged from the solid state theory at the point of fundamental approach about 100 years ago: the nature of collective excitations in liquids was not known at the time when this divergence took place. With the liquid theory operating in terms of collective excitations as in the solid state theory, we are now in a strong position to understand liquids.

This book reviews this research, starting with the early work by Sommerfeld and Brillouin and ending with recent independent verifications of the liquid theory. I will follow the variation of the phase space in liquids in a wide range of parameters on the phase diagram, from low-temperature liquids to high-temperature supercritical fluids. I will then come back to low-temperature viscous liquids approaching liquid-glass transition.

The liquid theory and its independent verifications discussed in this book focus on real liquids and their experimental properties rather than on model systems. This importantly differentiates this book from others.

I will also show how developments in liquid theory resulted in new unexpected insights. For example, the variation of the phase space available to phonons is related to liquid viscosity which quantifies the ability to flow. Viscosity has a minimum related to the crossover of particles dynamics from liquidlike to gaslike. It turns out that this minimum is governed by fundamental physical constants including $\hbar$. This, in turn, shows that water and water-based life are well attuned to fundamental physical constants including the degree of quantumness of the physical world set by $\hbar$. Discussing the implications of this adds another layer to the anthropic principle.

Tabor \index{Tabor} calls liquids ``neglected step-child of physical scientists'' and ``Cinderella of modern physics'' as compared to solids and gases \cite{tabor}. Although this observation was made nearly 30 years ago, Tabor would have reached the same conclusion regarding liquid thermodynamics on the basis of more recent textbooks (as discussed in the Introduction, liquid textbooks mostly focus on liquid structure and dynamics and do not discuss most basic thermodynamic properties such as liquid energy \index{energy} and heat capacity). \index{heat capacity} An important aim of this book is to make liquids a full family member on par with the other two states of matter, if only more sophisticated due to the liquid ability to sustain a variable phase space.

The selection of topics in this book was helpfully aided and narrowed down by adopting a well-established approach in physics where an interacting system is fundamentally understood on the basis of its excitations \cite{landaustat,landaustat1}. Consequently, a large part of this book discusses collective excitations \index{collective excitations} in liquids and their relation to basic liquid properties throughout the history of liquid research. This shows how earlier and more recent ideas physically link to each other and in ways not previously considered.

I perceive that there will be several groups of readers benefiting from better understanding liquids. In his recent review, Chen \cite{chen-review} observes that our current understanding of thermophysical properties of liquids is very unsatisfactory, adding on a more optimistic note that progress will be helped by what has been done earlier. This book therefore reaches out to scientists at any stage of their career who are interested in the states of matter and a history of a long-standing problem of understanding liquids theoretically. The second group are researchers and graduate students working in the area of liquids and related areas such as soft condensed matter \index{condensed matter} physics and systems with strong dynamical disorder. The third group are lecturers looking to include liquids in the undergraduate and graduate courses such as statistical or condensed matter physics as well as students who can use this book as a reference. As observed by Granato \cite{granato},

\begin{quote}
``there is nothing said about [liquid specific heat] in the standard introductory textbooks, and little or nothing in advanced texts as well. In fact, there is little general awareness even of what the basic experimental facts to be explained are.''
\end{quote}

This observation has been recently shared and highlighted, emphasising the lack of undergraduate instruction in fluids \cite{prescod}. This book aims to fix this issue for the benefit of both lecturers and their students.

I am grateful to M. P. Allen, V. Brazhkin, B. Carr, G. Chen, L. Noirez, J. C. Phillips and U. Winderberg for discussion and comments, I. Shkoropad for illustrations and to my family for support.


\newpage

\section{Introduction}

I had a memorable library day trying to find an answer to a question that is simple to formulate: what is a theoretical value of energy \index{energy} and heat capacity \index{heat capacity} of a classical liquid? I looked through all textbooks dedicated to liquids as well as statistical physics and condensed matter textbooks in the Rayleigh Library at the Cavendish Laboratory in Cambridge. To my surprise, they had either very little or nothing to say about the matter. I then took a short walk to Cambridge University Library and did a thorough search there. This returned the same result.

My surprise quickly grew closer to an astonishment, for two reasons. First, the heat capacity \index{heat capacity} is one of the central properties in physics. Constant-volume heat capacity is the temperature derivative of the system energy, the foremost property in physics including statistical physics. Heat capacity informs us about the system's degrees of freedom \index{degrees of freedom} and regimes in which the system is in, classical or quantum. It is also a common indicator of phase transitions, their types and so on. Understanding energy \index{energy} and heat capacity of solids and gases is a central and fundamental part of theories of these two phases. Thermodynamic properties such as energy and heat capacity are also related to important kinetic and transport properties such as thermal conductivity. Not having this understanding in liquids, the third basic state of matter, is a glaring gap in our theories. This is especially so in view of enormous progress in condensed matter research in the last century.

The second reason for my surprise was that the textbooks did not mention the absence of discussion of liquid heat capacity as an issue. It is harder to solve a problem if we don't know it exists.

Around the same time, I met Professor Granato from the University of Illinois who shared my observations. He notes \cite{granato} that nothing is said about liquid specific heat \index{specific heat} in standard introductory textbooks, and little or nothing in advanced texts, adding that there is little general awareness of what the basic experimental facts to be explained are. This was reflected in his teaching: Granato recalls how he enjoyed teaching theories of specific heat of gases and solids but lived in constant fear that a student might ask ``how about liquids''? That question was never asked over many years of teaching involving about 10,000 students. Granato attributes this to an issue with our teaching practice where unsolved problems in physics are not sufficiently covered.

The available textbooks are mostly concerned with liquid structure and dynamics. They do not discuss most basic thermodynamic properties such as liquid energy \index{energy} and heat capacity \index{heat capacity} or explain whether the absence of this discussion is related to a fundamental theoretical problem. Textbooks where we might expect to find this discussion but don't, include those dedicated to liquids \cite{frenkel,fisher,kirkwoodbook,faber,ubbelohde,march,boonyip,marchtosi,tabor,egelstaff,faber1,balucani,hansen2,hansen1,evans,gallo} and related systems \cite{ziman,edwards,soft1,kob,soft,parisi}, advanced condensed matter \index{condensed matter} texts \cite{chaikin,fradkin,cohen,anderson,girvinyang} and statistical physics textbooks \cite{landaustat,landaustat1,toda,chandler,kubo,ma,kittel,kardar,reif,sethna} \footnote[2]{I will discuss two books by D. Wallace \index{Wallace} and J. Proctor \index{Proctor} separately.}.

This list has a notable outlier: the Statistical Physics textbook by Landau and Lifshitz \index{Landau and Lifshitz} \cite{landaustat}. Landau and Lifshitz discuss general thermodynamic properties of liquids and explain why they {\it can not} be calculated, contrary to solids and gases. This explanation is quoted in the Preface, together with a related quote from Pitaevskii discussing the absence of a small parameter in liquids. The absence of a small parameter implies that we can't use well-known tools such as quadratic forms \index{quadratic forms} based on the harmonic expansion in solids and perturbation theory \index{perturbation theory} in gases in order to understand liquids theoretically. Other authors made similar observations which we will meet throughout this book. For example, Tabor says that ``the liquid state raises a number of very difficult theoretical problems'' \cite{tabor}.

As set out by Landau, Lifshitz and Pitaevskii, the liquid problem is formidable. It could explain why liquid energy and heat capacity are not discussed in textbooks, although in my experience, shared by Granato, the lack of these discussions may be related to little awareness of experimental facts needed explanation, let alone to the fundamental problem of liquid theory.

The problem of liquids was well recognised in the area of molecular modelling and molecular dynamics simulations. \index{molecular dynamics simulations} It is interesting that a classic textbook on molecular dynamics simulations is entitled ``Computer Simulation of Liquids'' \cite{allentild}. This method can simulate solids, gases or any other system where an interatomic interaction is known, yet the textbook title involves liquids. As the book authors explain, an important motivation for computer simulations as such was the need to understand liquids because they are not amenable to theoretical understanding using approximate perturbative approaches or virial expansions. The authors note that for some liquid properties, it may not even be clear how to begin constructing an approximate theory in a reasonable way. Computer modelling was seen as a way to help fix this problem.

In view of enormous progress of condensed matter research, it is perhaps striking to realise that we do not have a basic understanding of liquids as the third state of matter and certainly not on par with solids and gases. This was one of the reasons I decided to look into this problem.

As its title suggests, this book follows the path from excitations in liquids to their thermodynamic properties. This general path is well-established in statistical physics: an interacting system has collective excitations, or collective modes. Thermodynamic properties are then calculated on the basis of these collective excitations. The solid state theory \index{solid state theory} is a celebrated case where the solid energy is calculated on the basis of phonons in the Einstein \index{Einstein theory of solids} or Debye model, \index{Debye model} with all important consequences that follow. In liquids, this path was largely unexplored in the past (with an exception of liquid helium where the quantum nature of the system, perhaps surprisingly, simplifies the theory). Till fairly recently, collective excitations in liquids were not understood well enough, and it was unclear whether and how they can be related to liquid thermodynamics on the basis of a quantitative theory.

As is often the case, a hard problem needs to be addressed from different perspectives, and the first aim of this book is to synthesise results coming from three different lines of enquiry: experiment, theory and molecular modelling. Some of these results are fairly new, whereas others date back nearly two centuries. I will show that liquids have a long and extraordinary history of research but this history has been fragmented and sometimes undervalued.

Once looked from a longer-term perspective, the history of collective excitations \index{collective excitations} in liquids reveals a fascinating story which involves physics luminaries and includes milestone contributions from Maxwell \index{Maxwell} in 1867, followed by Frenkel \index{Frenkel} and Landau. \index{Landau} A separate and largely unknown line of enquiry aiming to connect phonons in liquids and liquid thermodynamics involved the work of Sommerfeld \index{Sommerfeld} published in 1913, one year after the Debye's \index{Debye model} paper on heat capacity \index{heat capacity} of solids. This line of enquiry was later developed by Brillouin \index{Sommerfeld} and Wannier and Pirou\'e. \index{Wannier and Pirou\'e} When we get to the equation written (but not solved) by Frenkel \index{Frenkel} to describe collective excitations in liquids, we will see that this equation was introduced by Kirchhoff \index{Kirchhoff} in 1857 and discussed by Heaviside \index{Heaviside} and Poincar\'e. \index{Poincar\'e}

The second and most technical aim of this book is to connect excitations in liquids to their thermodynamic properties. In the process, a distinct liquid story will emerge, and this will achieve the third aim of this book: to put real liquids and understanding their experimental properties in spotlight. Tabor \index{Tabor} calls liquids ``neglected step-child of physical scientists'' and ``Cinderella of modern physics'' as compared to solids and gases \cite{tabor}. Although this observation was made nearly 30 years ago, Tabor would have reached the same conclusion today with regard to liquid thermodynamics on the basis of more recent textbooks listed earlier. The third aim of this book is therefore to make liquids a full family member on par with the other two states of matter, if only more sophisticated and refined due to their ability to combine oscillatory solidlike and diffusive gaslike components of particle motion. We will see how these two components endow the phase space for collective excitations \index{collective excitations} in liquids with an important property: its ability to change with temperature.

The focus of this book is on understanding real liquids and their experimental properties rather than on model systems. This importantly differentiates this book from others.

The experimental properties include the specific heat \index{specific heat} as an important indicator of the degrees of freedom \index{degrees of freedom} in the system and a quantity that can be directly measured experimentally and compared to a theory. I start with a brief account of experimental data we will be mostly concerned with in Chapter 3. It will become quickly apparent that common models used to understand liquids such as the van der Waals \index{Van der Waals model} and hard-sphere models \index{hard-sphere model} are irrelevant for understanding the specific heat of real liquids, bringing about the realisation that something quite different is needed. I will then recall our current understanding of collective modes in solids and gases and how these modes underly the theory of their thermodynamic properties in Chapters 4 and 5. An important observation here is that quadratic forms \index{quadratic forms} greatly simplify theoretical description. We particularly appreciate this simplification when we deal with liquids where we don't have them.

The small parameter \index{small parameter} problem discussed by Landau, Lifshitz and Pitaevskii \index{Landau and Lifshitz} above does not necessarily mean that liquids can not be understood in some other way. However in Chapter 6 we will see that a head-on first-principles description of liquids which we use in the solid state theory \index{solid state theory} is exponentially complex and therefore intractable.

An alternative approach to liquids is based on an integral involving interactions and correlation functions, and I review this approach in Chapter 7. This will show that earlier liquid theories and the solid state theory diverged at the point of a fundamental approach. Early theories  considered that the goal of statistical theory of liquids is to provide a relation between liquid thermodynamic properties and the liquid structure and intermolecular interactions. Working towards this goal involved ascertaining the analytical models for structure and interactions in liquids. Developing these models has then become the essence of earlier liquid theories. I will review general problems involved in this approach including the inevitable problems set out by Landau, Lifshitz and Pitaevskii in the Preface. I will then observe that these problems are absent in the solid state theory because this theory does not operate in terms of correlation functions and interactions and is based on phonons instead. This makes the solid state theory tractable, physically transparent and predictive. One of the main aims of this book is to show that liquids can be understood using a similar theory based on phonons.

In Chapter 8, I will show how Frenkel's \index{Frenkel} ideas broke new ground for understanding liquids. This notably involves the concept of liquid relaxation time and its implications for liquid dynamics and collective excitations. \index{collective excitations} This resulted in several important predictions, including the propagation of solidlike transverse waves \index{transverse wave} in liquids at high frequency. The main difference between liquids and solid glasses was thought to be resistance to shear stress, however Frenkel \index{Frenkel} proposed that this is not the case: liquids too can support shear stress, albeit at high frequency. Frenkel also proposed that the liquid-glass transition \index{glass transition} is a continuous, rather than a discontinuous, process. As we will see in Chapter 17, this continues to be the topic of current glass transition research. Although Frenkel broke new ground for understanding phonons in liquids, he did not seek to connect these phonons to key thermodynamic properties of liquids such as energy and heat capacity.

In Chapter 9, I will discuss the equation governing liquid transverse modes that Frenkel \index{Frenkel} wrote but, surprisingly, did not solve. Once simplified, this equation becomes the ``telegraph equation'' discussed by Kirchhoff, \index{Kirchhoff} Heaviside \index{Heaviside} and Poincar\'e. \index{Poincar\'e} Surprisingly again, in view of its wide use, the telegraph equation \index{telegraph equation} was not fully explored in terms of different dispersion relations it supports. I will show how the solution gives rise to an important property of collective transverse modes in liquids: these modes exist above the threshold wavevector only and correspond to {\it gapped} \index{momentum gap} momentum states. In terms of propagating waves, the gap also exists in the frequency domain. The gap increases with temperature, and this brings about the key property of liquids: the volume of the phase space available to collective excitations \index{collective excitations} {\it reduces} with temperature. This is in striking difference with solids where this phase space is fixed. I will discuss the evidence for this reduction of the phase space \index{phase space, reduction} on the basis of experimental and modelling data.

I will observe that the hydrodynamic description was historically adopted as a starting point of discussing liquids and their viscoelastic properties. \index{viscoelasticity} In this approach, hydrodynamic equations are modified to account for solidlike response. I will show that this is related to our subjective perception rather than liquid physics and that both hydrodynamic and solidlike elastic theory are equally legitimate starting points of liquid description, both resulting in the telegraph equation. \index{telegraph equation} In this sense, liquids have a symmetry of their description where the ``elastoviscosity'' term is as justified as the commonly used term ``viscoelasticity'' to discuss liquid properties.

As mentioned earlier in this Introduction, widely-used molecular dynamics simulations were stimulated by the need to understand liquids. This was because liquids were not thought to be amenable to theory. In Chapter 10, I will review the molecular dynamics (MD) simulations method and focus on two sets of MD data related to the main topic of this book: thermodynamic properties and $c_v$ in particular and collective excitations.

Armed with preceding results, I will discuss the calculation of liquid energy and specific heat on the basis of propagating phonons in Chapter 11. I will demonstrate the advantage of this approach over that based on interatomic potentials and correlation functions. Theoretical results will be compared to a wide range of experimental data universally showing the decrease of the liquid specific heat with temperature. This decrease is related to the reduction of the volume of the phase space \index{phase space, reduction} available for collective excitations due to the progressively smaller number of transverse phonons at high temperature. I will also discuss independent verifications of this theory.

In the following Chapter 12, I will examine the relation between collective excitations in quantum liquid $^4$He and its specific heat as discussed by Landau \index{Landau} and Migdal. \index{Migdal} I will also discuss $^3$He where, differently from other systems, low-temperature thermodynamic properties are governed by localised quasiparticles rather than collective modes.

In Chapter 13, I will recall the earlier Tabor's \index{Tabor} discussion of what constitutes the ``{\it Sui Generis}'' approach to liquids. In the light of preceding results and our current understanding, the key to Sui Generis in liquids is to observe that these are systems in a {\it mixed} dynamical state involving both oscillations and diffusion of particles, in contrast to pure dynamical states in gases and solids where particle motion is purely diffusive and oscillatory. The balance between oscillatory and diffusive motion shifts with temperature (pressure), representing the sophistication of liquids as compared to solids and gases. This shift is related to the variation of the volume of phase space \index{phase space, reduction} available to collective excitations \index{collective excitations} mentioned earlier. This, in turn, gives the key to understanding liquid thermodynamics and the universal decrease of liquid specific heat \index{specific heat} with temperature in particular.

In Chapter 14, I will review the history of research aiming to link liquid thermodynamic properties to liquid collective excitations. Involving physics luminaries, this history dates back over a century. Quite remarkably, the first attempt at the theory of liquid thermodynamics based on phonons was done by Sommerfeld \index{Sommerfeld} only one year after the Debye \index{Debye model} theory of solids and 5 years after the Einstein's \index{Einstein theory of solids} theory were published. Whereas the Debye theory has become part of every textbook where solids are mentioned, we had to wait for about a century until we started understanding phonons in liquids well enough to be able to connect them to liquid thermodynamic properties. The Sommerfeld's \index{Sommerfeld} line of enquiry was taken up by Brillouin, \index{Brillouin} who reconciled his theory of liquids and experimental data by making a fascinating proposal that liquids may consist of small crystallites. Wannier and Pirou\'e \index{Wannier and Pirou\'e} later expanded on this work, as did other authors.

I will follow the link between excitations in liquids and their thermodynamics in a wide range of pressure and temperature on the phase diagram, \index{phase diagram} from high-temperature supercritical fluids \index{supercritical fluid} \index{supercritical fluid}to subcritical \index{subcritical liquid} liquids and to low-temperature viscous systems approaching the glass transition. \index{glass transition} In Chapter 15, I will examine the reduction of the phase space available to collective excitations above the critical point. \index{critical point} Above the Frenkel line \index{Frenkel line} corresponding to the disappearance of the oscillatory component of particle motion and two transverse modes, this reduction is due to progressive disappearance of the remaining longitudinal mode. \index{longitudinal wave} In view of increasing deployment of supercritical fluids in environmental, cleaning and extracting applications, this discussion is of environmental and industrial relevance, as Chapter 15 will show.

The phase space available to collective excitations \index{collective excitations} depends on the Frenkel hopping frequency \index{hopping frequency} which, in turn, depends on viscosity. \index{viscosity} The temperature dependence of viscosity shows universal minima, and the minima themselves are related to the dynamical crossover of particle dynamics discussed in Chapter 15. This leads us to Chapter 16 where I show that viscosity {\it minimum} is a universal quantity for each liquid and is fixed by fundamental physical constants, an interesting result in view that viscosity is strongly system- and temperature-dependent. I will show how this provides an answer to the question asked by Purcell \index{Purcell} and considered by Weisskopf \index{Weisskopf} in the 1970s, namely why viscosity never drops below a certain value comparable to that of water? \index{water}

The minimal viscosity \index{minimal viscosity} \index{viscosity, minimal} turns out to depend on $\hbar$ and hence be a quantum property. This has immediate consequences for water viscosity and essential processes in living organisms and cells. Water \index{water} and life appear to be well attuned to the degree of quantumness of the physical world and other fundamental physical constants. This adds another layer to the anthropic principle. \index{anthropic principle}

Going to another extreme of large viscosity at low temperature brings us to the realm of viscous liquids discussed in Chapter 17. Here, the link between excitations and system thermodynamics becomes fairly simple because the reduction of the phase space \index{phase space, reduction} available to collective excitations can be safely ignored. As a result, liquid energy and specific heat to a very good approximation are given by $3N$ phonons as in solids. Together with the dynamical disappearance of the viscous response at the glass transition \index{glass transition} temperature, this explains the observed jump of liquid specific heat at the glass transition temperature which logarithmically increases with the quench rate.

In Chapter 18, I will discuss Mott's recollection of the Frenkel work. Mott says that Frenkel was interested in real systems and what is really happening in those systems. This prompts a discussion of what we aim to achieve by a physical theory and what this theory should do.

The last two Chapters 19 and 20 explore how the insights we learned from the liquid theory can be used in other areas. Chapter 19 discusses how spin glasses can be understood on the basis of spin waves \index{spin waves} as is done in the theory of ferromagnets and antiferromagnets and similarly to how structural glasses are understood on the basis of phonons. In Chapter 20, I discuss the problem of strongly-coupled field theories and make an analogy with strongly-interacting liquids.

I have not discussed all sources I have consulted. The selection of topics was helpfully aided and narrowed down by the guiding principle I used in writing this book. This principle is based on a well-established approach in physics where an interacting system is fundamentally understood on the basis of its excitations \cite{landaustat,landaustat1}. I have also taken a cautious attitude towards proposals that have not yet shown themselves as well-founded or constructive (I briefly touch upon this issue in Chapter 18). I may have missed subtle or more important points, and for this reason I recommend that an inquisitive reader consult references for details and look elsewhere. This is particularly so in view that this book includes the results of my own group and collaborators. But where I have really made a difference is in writing this overview as more than a collection of ideas, concepts and models related to liquids. By focusing on collective excitations \index{collective excitations} in liquids and basic liquid properties, I have shown that earlier and more recent research and ideas physically link to each other and in ways not previously considered.

I end several chapters with questions and proposals for future research to either complement the already existing results or pursue a new line of enquiry.



\newpage

\section{Experimental heat capacity}
\label{intro}

\begin{figure}
\begin{center}
{\scalebox{0.55}{\includegraphics{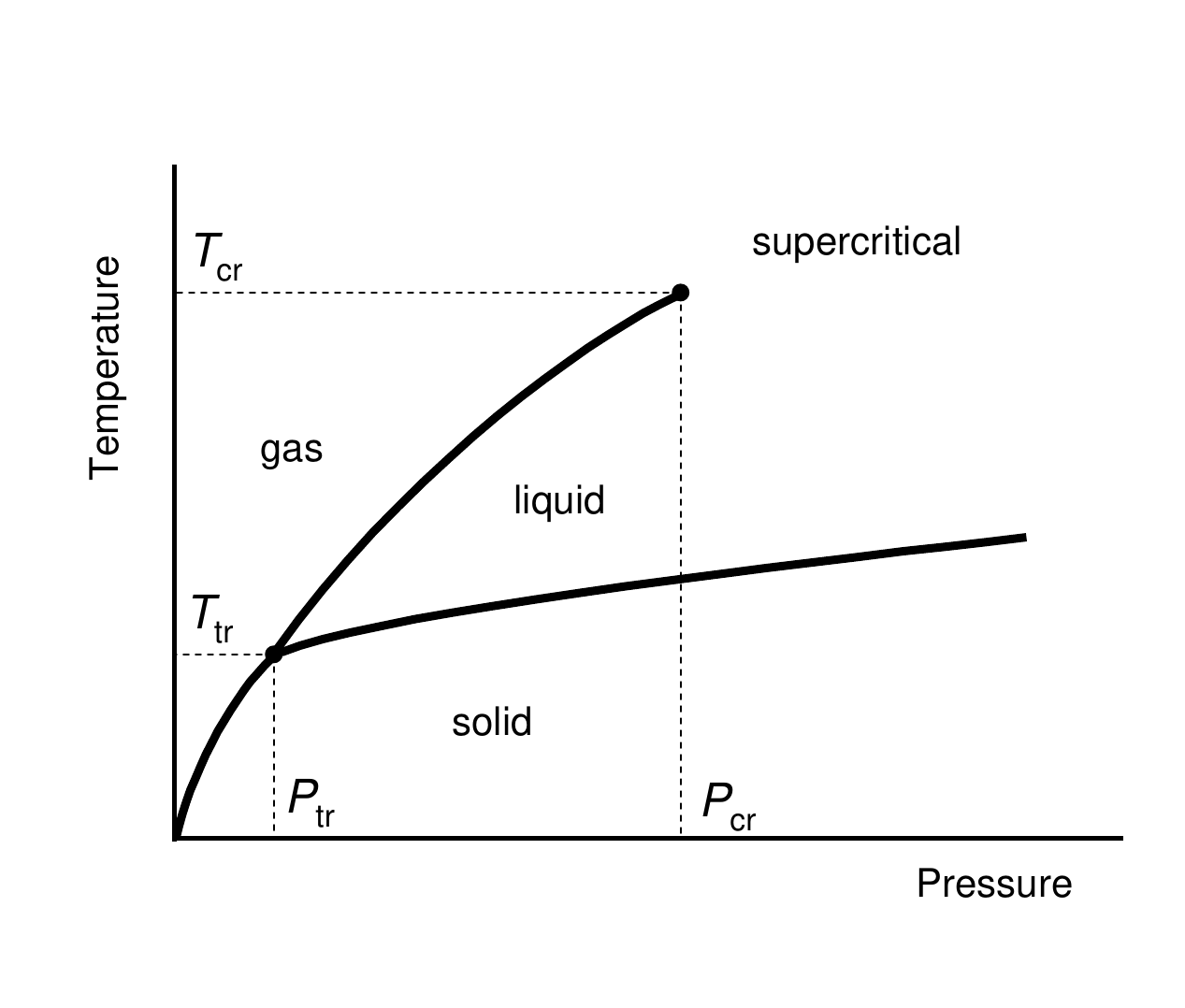}}}
\end{center}
\caption{Phase diagram of matter with triple and critical points shown.}
\label{phase}
\end{figure}

Our understanding of the basic states of matter is illustrated by the phase diagram \index{phase diagram} shown in Figure \ref{phase}. Solids are located in the low-temperature and high-pressure part of the phase diagram. Gases are in the opposite high-temperature and low-pressure part. Liquids are sandwiched between solids and gases: the liquid state starts above the triple point \index{triple point} \index{critical point} and is bound by melting and boiling lines.

As discussed in the Preface and Introduction, there is large and striking gap between theoretical understanding of the three states of matter. Theories of solids and gases are well developed. In stark contrast, there is no theoretical understanding of the most basic liquid properties such as energy \index{energy} and specific heat. \index{specific heat} This book seeks to reduce this gap and is concerned with fundamental understanding of liquids, with the emphasis on real liquids and experimental data rather than on model systems.

We will adopt a generic approach of statistical physics where properties of an interacting system are fundamentally related to excitations. The solid state theory \index{solid state theory} is one notable example where this relation is used in many important ways. Our broad goal in this book is to ascertain the nature of collective excitations \index{collective excitations} in liquids and then relate them to liquid thermodynamic properties.

Returning briefly to the phase diagram \index{phase diagram} in Figure \ref{phase}, we see that the boiling line ends at the critical point \index{critical point} which conditionally marks the beginning of the supercritical state. \index{supercritical state} A large part of this book deals with subcritical liquids. \index{subcritical liquid} Towards the end if this book, I will also discuss how the insights from the theory of subcritical liquids can be used to understand the supercritical state and its dynamical and thermodynamic properties.

\begin{figure}
\begin{center}
{\scalebox{0.35}{\includegraphics{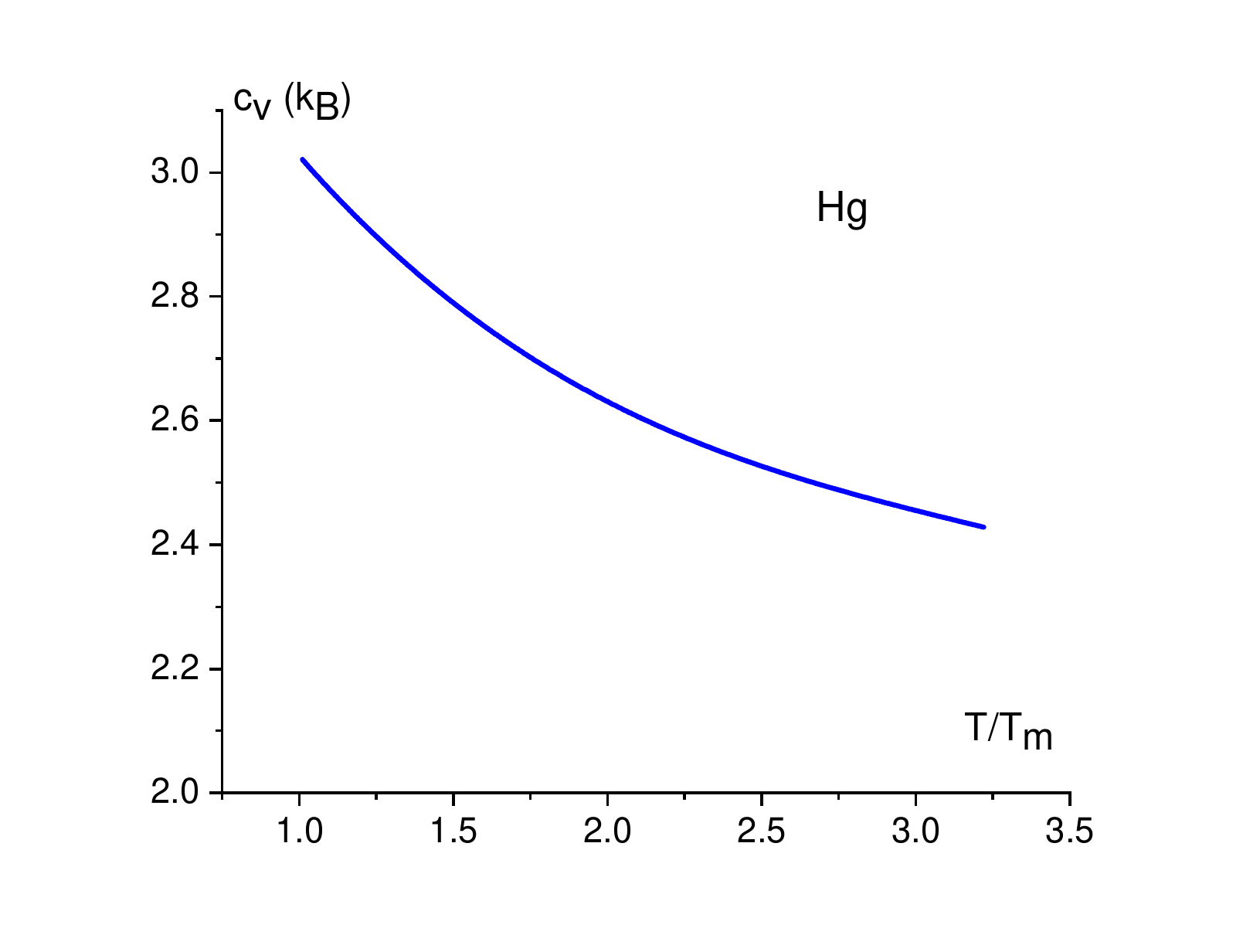}}}
{\scalebox{0.35}{\includegraphics{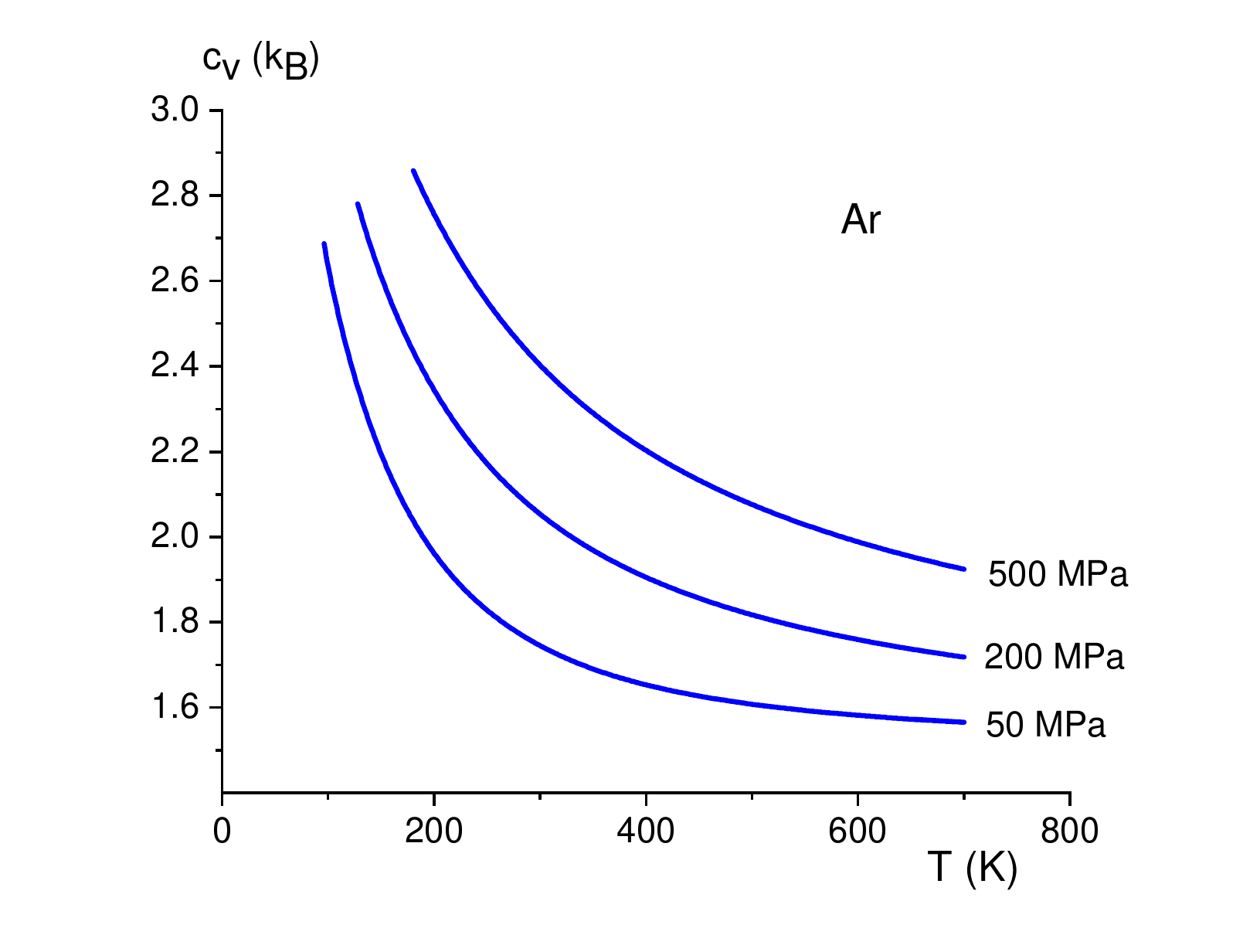}}}
\end{center}
\caption{Top: experimental specific heat \index{specific heat} $c_v$ of liquid mercury with the electronic contribution subtracted as a function of reduced temperature, where $T_m$ is the melting temperature \cite{grimvallcv,wallacecv}. Bottom: experimental $c_v$ of Ar at different pressures \cite{nist}.
}
\label{mercury}
\end{figure}

One of my talks in Cambridge was interrupted by a keen theoretician, who asked to show the experimental data early on. I later thought this is a good idea because it quickly tells us what we are dealing with. Figure \ref{mercury} shows the experimental specific heats \index{specific heat} of liquid mercury \index{mercury} and argon. \index{argon} These two liquids represent different liquid types: metallic and noble. \index{metallic liquid} \index{noble liquid}

Here and below, we adopt the notations of Ref. \cite{landaustat} and set $k_{\rm B}=1$ everywhere.

$c_v$ of liquid mercury, \index{mercury} with the electronic contribution subtracted, is shown in Figure \ref{mercury} in the temperature range where mercury is subcritical. We also show $c_v$ of liquid argon. \index{argon} Argon is well described by a simple Lennard-Jones potential. For this reason, liquid argon has become a common case study in numerous theoretical approaches and molecular dynamics simulations. We show $c_v$ of Ar at three different pressures. The lowest pressure exceeds the critical pressure $P_c\approx 4.9$ MPa by about ten times in order to expand temperature range where Ar exists as a fluid (at room pressure this range is quite narrow) and to avoid critical and near-critical anomalies affecting $c_v$. This book considers generic behavior of $c_v$ unaffected by critical and near-critical anomalies. We will discuss the effects related to the critical point in Section 15B and the supercritical state \index{supercritical state} in Chapter 15.

We observe that liquid $c_v$ of liquid mercury is close to 3 at low temperature and tends to about 2 at high. A similar trend is seen in liquid Ar: $c_v$ starts just below 3 at low temperature and decreases to 2 and below.

We also observe that at a fixed temperature, $c_v$ increases with pressure. In Ar, viscosity \index{viscosity} increases with pressure and, therefore, larger viscosity is related to larger $c_v$. It is useful to keep this in mind for now. We will revisit pressure dependence of $c_v$ in Section 11E.

We will later see that the decrease of $c_v$ from 3 to 2 and below to its limiting ideal-gas value of $\frac{3}{2}$ is a universal behavior seen in other metallic, noble, molecular and hydrogen-bonded liquids. A large part of this book is about understanding this behavior theoretically.

There are three further quick observations from Fig. \ref{mercury}. First, the decrease of $c_v$ is different from what is seen in solids where $c_v$ either increases with temperature due to phonon excitations or stays constant in the classical harmonic regime.

Second, $c_v$ at low temperature is close to the Dulong-Petit \index{Dulong-Petit} value of 3. This result, revisited later in this section on the basis of the Wallace \index{Wallace} plot, immediately suggests that the physics of liquids close to melting is governed by $3N$ phonons \index{phonon} ($N$ is the number of particles) as is the case in solids. Another possibility is that some other mechanism operates in liquids which nevertheless gives the Dulong-Petit result $c_v=3$, or that this other mechanism and phonons work together to give $c_v=3$. This is unlikely on general physical grounds since it would involve fine tuning to give the same universal low-temperature value $c_v=3$ in all liquids. We will later see that no mechanism other than phonons is needed to account for the experimental value of specific heat. We will find that this assertion requires several important sets of results from theory, experiment and modelling.

The third and for now final observation from Figure \ref{mercury} is that common models widely used to discuss liquids are irrelevant for understanding the energy \index{energy} and heat capacity \index{heat capacity} of real liquids. These models are notable workhorses of liquid physics and include the widely discussed van der Waals model \index{Van der Waals model} \cite{hansen2} and the hard-sphere model \index{hard-sphere model} (see, e.g., Ref. \cite{barkerhenderson,parisi,parisihard}). This may come as perhaps surprising to the reader, in view of the wide discussion of these models in the literature and textbooks. For this reason, it is worth expanding upon this point.

Our first example is the van der Waals model (vdW) widely used to discuss liquid-related effects, including the liquid-gas transitions, critical points, compressibility, liquid's ability to withstand negative pressure and as a starting point of liquid theories involving perturbation methods \index{perturbation theory} \cite{hansen2}. The vdW equation of state reads

\begin{equation}
\left(P+\frac{N^2a}{V^2}\right)\left(V-Nb\right)=NT
\label{vdw}
\end{equation}

\noindent where $P$, $V$, $N$ and $T$ are pressure, volume, number of atoms and temperature, and $a$ and $b$ are constants related to the pair interaction potential and atomic size.

Eq. \eqref{vdw} corresponds to the free energy \cite{landaustat}: \index{free energy}

\begin{equation}
F=F_{id}-NT\ln\left(1-\frac{Nb}{V}\right)-\frac{N^2a}{V}
\label{vdwfree}
\end{equation}

\noindent where $F_{id}$ is the free energy of the ideal gas.

Eq. \eqref{vdwfree} follows from (a) assuming a dilute system with small density and corresponding to the first two terms in the virial expansion in powers $\frac{N}{V}$, (b) calculating the temperature-dependent term of the second virial coefficient in the high-temperature approximation $\frac{U_0}{T}\ll 1$, where $U_0$ is the cohesion energy and (c) interpolating between the liquid and gas state and imposing a finite compressibility in the liquid where the liquid volume $V$ should always be larger than $Nb$ in \eqref{vdwfree} (or else the argument of the logarithm becomes negative) \cite{landaustat}.

The entropy $S=-\left(\frac{\partial F}{\partial T}\right)_V$ depends on temperature through the first ideal gas term only:

\begin{equation}
S=S_{id}+N\ln\left(1-\frac{Nb}{V}\right)
\end{equation}

The energy \index{energy} $E=F+TS$ is

\begin{equation}
E=E_{id}-\frac{N^2a}{V}
\end{equation}

\noindent resulting in the specific heat \index{specific heat} of the ideal gas \cite{landaustat}:

\begin{equation}
c_v=\frac{3}{2}
\label{ideal}
\end{equation}

\noindent in stark contrast to the experimental data in Figure \ref{mercury}.

We see that the vdW model describes an ideal gas from the point of view of system's specific heat. This model is a good example of how the virial expansion and high-temperature approximations, while accounting for some liquid properties and introducing interactions in the system, are nevertheless unable to alter $c_v$ and the associated number of degrees of freedom \index{degrees of freedom} in the ideal gas reference system.

We might think that $c_v=\frac{3}{2}$ of the vdW model is related to the system being dilute. However, the vdW equation \eqref{vdw} is often derived and discussed by substituting $V$ by $V-Nb$ without assuming that $Nb$ is small compared to $V$ \cite{vdwcondmat}. Hence a system can have large solid-like concentration comparable to the inverse volume per atom and still have $c_v=\frac{3}{2}$. This brings us to the second workhorse of liquid physics, the hard-sphere model.

The hard-sphere model \index{hard-sphere model} consists of spheres of fixed volume set by an infinite interaction potential at distances shorter than the sphere radius and zero potential outside. This model has been widely used to account for the structure of liquids, giving rise to the approaches where short-range repulsion is dominant in governing liquid physics. We will review these theories in Chapter 7 in the context of perturbation approaches to liquids. The hard-sphere system has also been used to discuss liquid-solid transitions, equation of state, melting, communal entropy and so on (see, for example, References \cite{ziman,march,hansen2,barkerhenderson,parisi,parisihard}). This included the work of pioneers of computer simulations such as Alder, Wainwright and Hoover \cite{alder1,alder11,alder2,alder3}, laying the foundation for molecular dynamics simulations. \index{molecular dynamics simulations} We discuss these simulations in Chapter 10.

The specific heat \index{specific heat} of the hard-sphere system \index{hard-sphere model} is that of the ideal gas: $c_v=\frac{3}{2}$ as is the case of the vdW system in Eq. \eqref{ideal}. One easy way to understand this is to consider the hard-sphere system as the large $n$ limit of the interaction $U(r)\propto\frac{1}{r^n}$. Molecular dynamics simulations \index{molecular dynamics simulations} show that at large $n$, the ratio between the potential and kinetic energy becomes small, corresponding to the onset of the ideal gas behavior \cite{brazhkin2012}. This happens because harsh repulsion results in a particle spending an increasingly large proportion of its time moving outside the potential range, resulting in the average potential energy becoming small. With only kinetic energy remaining, the specific heat becomes equal to that of the ideal gas.

\begin{figure}
\begin{center}
{\scalebox{0.4}{\includegraphics{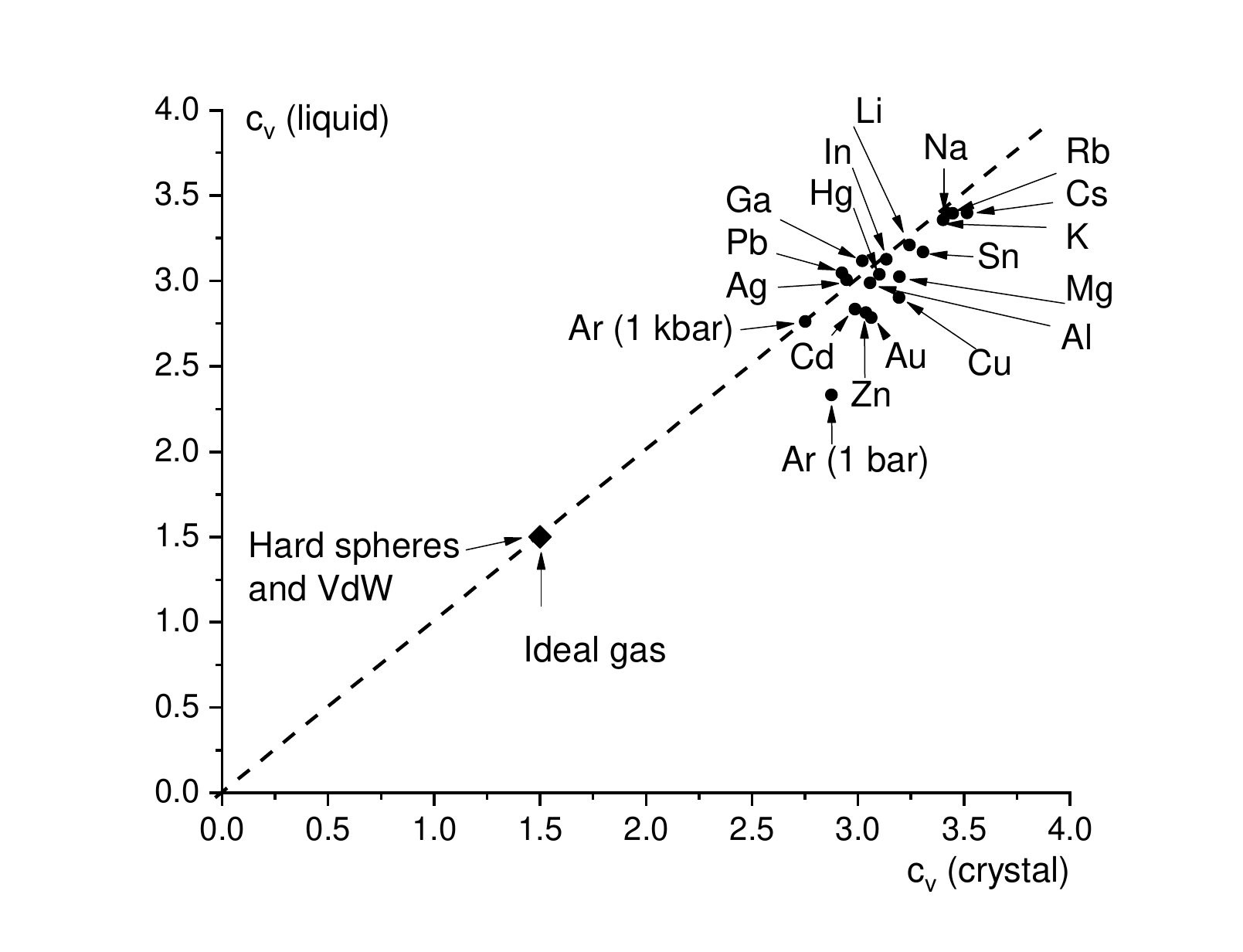}}}
\end{center}
\caption{The Wallace plot \index{Wallace plot} showing experimental $c_v$ of 18 liquids at the melting point against $c_v$ of crystals at the melting point. For metals, the electronic contribution if subtracted from the experimental data. The dashed line represents equality between liquid and crystal values. The ideal gas result $c_v=\frac{3}{2}$ for the hard-sphere model \index{hard-sphere model} is shown. The plot is adapted from References \cite{wallacecv,wallacebook} with permission from the American Physical Society. $c_v$ for the van der Waals \index{Van der Waals model} and the ideal gas systems are added to the Wallace plot.
}
\label{wallace}
\end{figure}

The irrelevance of the hard-sphere models for understanding liquid energy \index{energy} and $c_v$ is further illustrated in the Wallace \index{Wallace plot} plot. Wallace \cite{wallacecv,wallacebook} \index{Wallace} constructs a striking plot illustrating the similarity of experimental $c_v$ of liquids and their parent crystals for 18 systems shown in Figure \ref{wallace}. We have already noted this similarity earlier in this section. We observe that similarly to the vdW model, the hard-sphere model \index{hard-sphere model} is unable to change the number of degrees of freedom \index{degrees of freedom} from the ideal-gas value and to account for the experimental data.

Figure \ref{wallace} importantly shows that $c_v$ of real liquids is about twice as large as $c_v$ of the hard-sphere \index{hard-sphere model} and the vdW models where $c_v=\frac{3}{2}$ as in the ideal gas.

This large discrepancy also applies to liquid theories which go beyond the hard-sphere and vdW models while using these models as a starting point of a perturbation theory. \index{perturbation theory} A perturbation theory is tractable and meaningful in the sense of convergence only if terms in the series are small. Hence, adopting $c_v=\frac{3}{2}$ in the vdW or hard-sphere system as a starting point of the perturbation theory will not account for the experimental data in Figures \ref{mercury} and \ref{wallace} where $c_v=3$ at low temperature: to be consistent with experiments, the perturbation terms should account for another $\frac{3}{2}$ which is not small compared to the zeroth starting term. For this reason, perturbation methods are unable to describe the experimental $c_v$ shown in Figures \ref{mercury}-\ref{wallace} or explain the physical origin of behavior seen in these Figures. We will revisit this point in Section 7 where we will discuss liquid theories based on the hard-sphere \index{hard-sphere model} and vdW \index{Van der Waals model} models.

Our discussion of common liquid models such as the vdW and hard-sphere models sets us up for realisation that something quite different is needed to understand the experimental behavior of real liquids. This realisation is constructive and invites us think about what the starting point of liquid theory should be.

\newpage

\bibliographystyle{apsrev4-1}

\newpage
\begin{theindex}

  \item activation energy, 55
  \item anharmonicity, 38, 39, 41, 53, 90, 106, 144, 173
  \item anthropic principle, 15, 162, 163
  \item anyons, 121
  \item argon, 17, 138

  \indexspace

  \item bifurcation, 42, 44, 61
  \item Bloch law, 182
  \item Bogoliubov approximation, 31, 113
  \item Bohr, 104
  \item Bohr radius, 150
  \item Born, 127, 135
  \item Born and Green, 45, 47
  \item Bose liquid, 113, 147
  \item Boson peak, 88, 152
  \item Brillouin, 14, 52, 82, 125--127, 135, 138
  \item bulk modulus, 30, 64, 73, 83, 172

  \indexspace

  \item C transition, 137, 138
  \item collective excitations, 8, 11--16, 21, 26, 35, 45, 50, 52, 55,
		61, 75, 82, 89--94, 98, 113, 125, 140, 141, 166, 180
  \item condensed matter, 8, 10, 28, 53, 54, 75, 149, 151, 163
  \item continuum approximation, 21, 28, 29, 32, 58, 64, 79, 90, 141
  \item correlation function, 45, 48, 50, 52
  \item critical point, 14, 16, 125, 130, 131, 133, 136, 149, 157
  \item current correlation function, 75, 79--81, 94, 102

  \indexspace

  \item de Broglie wavelength, 119, 120
  \item Debye frequency, 37, 58, 100, 101, 103, 108, 143, 149, 182
  \item Debye model, 11, 14, 37, 50, 52, 66, 91, 100, 101, 105, 112,
		125--127, 180
  \item Debye vibration period, 138, 144, 150, 166, 174
  \item degrees of freedom, 10, 12, 19, 20, 112, 126, 177
  \item density of states, 25, 36, 37, 72, 90, 91, 100, 108, 113, 143,
		182
  \item dimensional analysis, 152
  \item dispersion relation, 25, 26, 32, 69, 70, 73, 74, 86, 88, 94,
		113
  \item Dulong-Petit, 17, 36, 38, 53, 91, 102, 107, 129, 145, 180

  \indexspace

  \item Einstein theory of solids, 11, 14, 35, 50, 52, 125, 127
  \item elasticity, 21, 28, 29, 61, 64, 87
  \item elasticity length, 70, 137, 138, 140, 145, 176
  \item elastoviscosity, 78
  \item energy, 8, 10, 16, 17, 19--24, 36, 37, 45, 98, 101, 130, 143
  \item energy gap, 64, 71
  \item entropy, 109, 147, 150, 168--170
  \item exponential complexity, 44, 55, 61

  \indexspace

  \item Faber, 129
  \item fast sound, 83, 118
  \item Fermi liquid, 116--118
  \item Fisher-Widom line, 136
  \item free energy, 19, 28, 35, 47, 107, 129, 181
  \item Frenkel, 11, 13, 21, 54--56, 58--61, 64--68, 171, 175
  \item Frenkel line, 14, 123, 130, 132, 135--138, 140, 146, 174
  \item frequency gap, 71, 72, 98
  \item fundamental constants, 160, 161, 164
  \item fundamental frequency, 176
  \item fundamental viscosity, 152

  \indexspace

  \item gapped energy state, 70
  \item gapped momentum state, 65, 71, 73, 74, 86, 94
  \item gas giants, 137
  \item generalised hydrodynamics, 61, 76, 79
  \item glass transition, 13--15, 56, 66, 67, 165, 168--171, 173,
		175, 176, 180--183
  \item Gr\"{u}neisen approximation, 40, 51, 106

  \indexspace

  \item hard-sphere model, 12, 17, 20, 21, 133
  \item harmonic oscillator, 23, 68
  \item harmonic solid, 23, 35, 39, 102, 167
  \item heat capacity, 8, 10, 11, 17, 21, 38, 98
  \item Heaviside, 12, 13, 69
  \item high-frequency response, 58, 60, 72
  \item honey, 77, 165, 168
  \item hopping frequency, 14, 55, 58, 70, 75, 98, 123
  \item hydrodynamics, 21, 28, 61, 64, 77

  \indexspace

  \item inversion point, 138, 145

  \indexspace

  \item Kauzmann temperature, 176
  \item Kirchhoff, 12, 13, 69

  \indexspace

  \item Landau, 11, 13, 67, 117, 125
  \item Landau and Lifshitz, 6, 10, 12, 22, 47--49, 51, 105
  \item life, 160, 162, 163
  \item lifetime, 39, 51, 69, 71, 90, 91
  \item liquid $^3$He, 116, 155
  \item liquid $^4$He, 113, 155, 156
  \item liquid relaxation time, 55, 58, 60, 64, 67, 80, 89, 111, 120,
		123, 127, 130, 135, 147, 165, 166, 185
  \item longitudinal wave, 14, 25, 26, 28, 30, 32, 36, 58, 64, 83, 98,
		115, 140, 141, 143

  \indexspace

  \item Maxwell, 11, 33, 41, 58--60, 69, 128, 149, 172
  \item mean free path, 117, 130, 138, 141, 147, 149
  \item mercury, 17, 60, 102, 165
  \item metallic liquid, 17, 103, 109, 138
  \item Migdal, 13, 114, 125
  \item minimal viscosity, 14, 147, 148, 151
  \item molecular dynamics simulations, 11, 20, 50, 57, 92, 102, 133,
		138
  \item molecular liquid, 103, 109, 112, 138, 157
  \item momentum gap, 13, 34, 65, 69--72, 81, 82, 85, 87, 88, 93, 94,
		105
  \item Mott, 55, 65, 178, 179

  \indexspace

  \item Navier-Stokes equation, 31, 60, 64, 68, 76, 77, 79, 161
  \item neutron scattering, 26, 82, 83, 113, 135, 181
  \item NMR, 181
  \item noble liquid, 17, 103, 109, 138, 157
  \item nonlinearity, 35, 41, 42, 61, 89
  \item normal modes, 25, 75

  \indexspace

  \item pair distribution function, 45, 46, 135, 136
  \item partition function, 35, 106
  \item perturbation theory, 11, 18, 20, 38, 39, 48, 185
  \item phase diagram, 14, 16, 131, 132, 134, 136, 140, 145, 157, 177
  \item phase space, reduction, 13--15, 51, 75, 94, 98, 102, 103, 105,
		110, 141, 146, 166, 183
  \item phonon, 17, 21, 26, 35, 38, 51, 55, 75, 82, 103, 113, 127, 138,
		144, 180
  \item phonon dispersion, 26, 52, 66, 83
  \item Pitaevskii, 6, 47, 48, 51, 105
  \item plane waves, 21, 25, 26, 29, 32, 74, 90, 185
  \item plasma, 85, 86
  \item Poincar\'e, 12, 13, 68, 69
  \item Proctor, 10, 111, 129
  \item propagation length, 63, 64, 70, 119, 176
  \item Purcell, 14, 148, 159, 160

  \indexspace

  \item quadratic forms, 11, 12, 22, 29, 35, 41, 51, 52, 105, 146, 185
  \item quantum field theory, 185
  \item quantum liquids, 113, 116, 119, 125
  \item quantum statistics, 31, 113, 119
  \item quark-gluon plasma, 147, 153

  \indexspace

  \item Raman scattering, 90, 135
  \item relaxation time, 55, 58, 60, 64, 67, 80, 89, 111, 117, 120, 123,
		127, 130, 135, 147, 165, 166, 185
  \item roton, 88, 113, 114
  \item Rydberg energy, 150

  \indexspace

  \item shear modulus, 55, 59, 60, 76, 127, 135, 145
  \item skin effect, 73
  \item small parameter, 6, 12, 41, 45, 51, 105, 185
  \item solid $^3$He, 120
  \item solid $^4$He, 120
  \item solid state theory, 11, 12, 16, 22, 35, 50, 52, 54, 125
  \item solubility, 137
  \item Sommerfeld, 11, 14, 52, 125--127
  \item sound wave, 28--30, 33, 34, 36, 113
  \item specific heat, 6, 10, 12, 14, 16--20, 36, 102, 139, 145, 167,
		170
  \item speed of sound, 26, 32, 37, 83, 86, 135, 147, 150
  \item spin dynamics, 183
  \item spin glasses, 180, 182
  \item spin waves, 15, 180, 181, 183
  \item stationary state, 44
  \item strong interactions, 6, 180, 185
  \item structure factor, 79, 83, 90, 93, 94
  \item subcritical liquid, 14, 16, 94, 129, 137, 140--142
  \item supercritical fluid, 14, 94, 95, 109, 133, 136, 140, 143, 144
  \item supercritical state, 16, 17, 34, 130, 131, 133, 136--138,
		140, 141, 148
  \item superfluid, 113, 116, 155
  \item susceptibility, 182

  \indexspace

  \item Tabor, 8, 12, 14, 123
  \item telegraph equation, 13, 68--70, 72, 76, 81, 82, 103
  \item thermal conductivity, 38, 69, 90, 135, 150, 157, 159
  \item thermal diffusivity, 157, 159
  \item thermal expansion, 38, 107, 131, 132, 144, 152, 169, 172, 173
  \item transverse wave, 13, 25, 33, 63, 71, 82, 83, 94, 98, 102, 182
  \item triple point, 16, 152

  \indexspace

  \item uncertainty relation, 154, 159

  \indexspace

  \item Van der Waals model, 12, 17, 20, 21, 47
  \item velocity autocorrelation function, 115, 133
  \item viscoelasticity, 13, 56, 60, 61, 64, 78, 172
  \item viscosity, 14, 17, 30, 58, 59, 64, 91, 111, 159, 165
  \item viscosity, fundamental, 153
  \item viscosity, generalized, 64, 76, 81
  \item viscosity, kinematic, 79, 80, 148
  \item viscosity, minimal, 14, 147--149, 151, 155
  \item viscosity, water, 163
  \item Vogel-Fulcher-Tammann equation, 175
  \item Vogel-Fulcher-Tammann equation, crossover, 175

  \indexspace

  \item Wallace, 10, 17, 20, 129
  \item Wallace plot, 20, 21
  \item Wannier and Pirou\'e, 11, 14, 128
  \item water, 14, 60, 137, 147, 159--162, 165
  \item Weisskopf, 14, 159
  \item Widom line, 131, 135, 136

  \indexspace

  \item X-ray scattering, 82, 83, 86, 135

\end{theindex}

\end{document}